# Modeling of 3D Printable Electrical Machineries Ferromagnetic Parts

**Shinthia Binte Eskender[1], Anupam Saha[2] and Shaikh Ishtiaque Ahmed[3]**

[1,2]Department of Materials Science and Engineering,
Khulna University of Engineering & Technology,
Khulna 9203, Bangladesh.
[3]Department of Industrial Engineering and Management,
Khulna University of Engineering & Technology,
Khulna 9203, Bangladesh.

E-mail: sinthiaoyshi@gmail.com

**ABSTRACT**

**The electrical machinery core is formed with a ferromagnetic material that offers high magnetic properties. As ferromagnetic materials have high relative magnetic permeability, they are important in the formation of electromagnetic device cores. Conventional subtractive and powder metallurgy methods for fabrication electrical machineries offer significant core losses and reduce magnetic flux density and magnetic permeability. With the advancement of technology, the limitation of the traditional process can be overcome by using the additive manufacturing process. Hence, this paper proposes a 3D printable model of two types of single-phase transformers, referred to as E-I shape and U-I shape transformers respectively. Possibilities of designing the electrical machinery part which has a ferromagnetic core are investigated. The efficiency of the transformers is evaluated in terms of magnetic flux density distribution and volumetric loss density based on the results of a large number of Finite element simulation methods under various operating situations on COMSOL. The performance of various ferromagnetic materials such as Soft Iron (Fe) and Ferrite ($Fe_2O_3$) on the transformer core is evaluated. This analysis reveals that if U-I shaped transformer can be made from 3D printing, it will be the best feasible structure for higher operating frequency.**

## 1. INTRODUCTION

Energy is required in every aspect of our life and its demand is increasing day by day. So, the conversion of energy is needed to fulfill the energy demand. Electric machineries, such as electric motors, electric generators, transformers, and other machines that use electromagnetic forces, are widely used for energy conversion. The cores of electric machines are traditionally made using subtractive techniques and the powder metallurgy process. In the powder metallurgy process, fine metal powders are compacted in suitable dies, and sintering is used to create metal parts. Subtractive manufacturing procedures remove material from a roughly shaped initial part to create the final shape and dimensions. Magnetic cores of electric machines have been typically manufactured by stacking electric steel laminations using subtractive processes [1].

Laminations made with traditional methods for extremely efficient electric motors still have significant losses. Eddy current loss is generally proportional to the thickness of the lamination in general [2]. Thinner laminations can reduce core loss. These are significant obstacles in the classic subtractive printing approach for electrical machines. Powder metallurgy has an impact on magnetic, electrical, and mechanical properties, in addition to limiting design form and size flexibility [3]. The non-magnetic gap between each particle in insulated iron powders applied in soft magnetic composites has a detrimental impact on magnetic permeability, strength, and magnetic flux density. The binder added to the powder is difficult since it drops the magnetic permeability and supreme flux density [4].

Additive manufacturing (AM) or 3D printing technologies are considered a crucial component of the industry 4.0 revolution, due to their improved capabilities above traditional manufacturing techniques. Fully 3D printed objects have been found to have several advantages over conventionally based manufacturing, such as it can be used to print thin-walled hollow parts with high electrical resistivity, complicated core designs may be realized, it has strong core loss reduction potential, and it can considerably lower fabrication costs [3]. With this advancement in technology, the limitation of the traditional subtractive and powder metallurgy process can be overcome by using the additive manufacturing process.

Ferromagnetic materials are those whose dipoles align with each other in the presence of an external magnetic field. Ferromagnetic materials have high relative magnetic

**International Conference on Mechanical, Manufacturing and Process Engineering (ICMMPE – 2022)**

Organizer: Faculty of Mechanical Engineering, Dhaka University of Engineering & Technology (DUET), Gazipur, Bangladesh

Website: https://icmmpeduet.com/permeability, which is crucial in the development of electromagnetic device cores that work with high magnetic flux for optimum performance.

In the previous study, Selective Laser melting (SLM), Binder jetting technology (BJT), Fused Deposition Modeling (FDM), and lamination object manufacturing (LOM) are the 3D printing technologies that have been used to build transformer, motor, and generator cores [2]–[6]. T. N. Lamichhane used FDM techniques for preparing ferromagnetic materials Machine Core [4], [7], [8]. In April 2021, [9] K. Onla worked on a 3D Multiphysics modeling of a three-phase transformer and proposed a three-dimensional Multiphysics model for temperature prediction of three types of three-phase transformers: Y-shape, three-columns, and five-columns transformers.

No researchers have shown the effect of 3D printable U-I and E-I shaped transformers on magnetic flux density and volumetric loss density based on FDM. In this study, U-I and E-I shaped transformer is designed by COMSOL Multiphysics. Then, the effect on magnetic flux density and volumetric loss density is simulated.

## 2. THEORIES OF ELECTRICAL MACHINERIES

An electrical machine converts mechanical energy into electrical energy or vice versa. Electric machineries, such as electric motors, electric generators, transformers, and other machines are widely used electric machineries in electrical engineering. Electrical motor application in the compression, drill machine, lathe machine, fans, refrigerator, crusher, pump, blower, and other machines. Radio, television, and telephone circuit used transformers, and for household purposes used distributed transformers [10]. Application of electrical generator-for Mining generator is used, a generator is used in RV camping, in construction site for power backup generator is used, and marine generator is used in boating. A transformer is a stationary device that converts alternating electricity from one voltage level to another (higher or lower), or from one voltage level to almost the same level, not changing its frequency. Mainly the motive of the transformer is to change the electrical voltage to a different value [1]. The transformer core's goal is to have a low-reluctance path whereby the greatest amount of flux of primary windings is transmitted through it and connected to the secondary winding. Transformers classified according to the lamination used core and shell-type are considered: U-I laminations, and E-I laminations. The shell-type transformer is a suitable decision for operating frequencies at the low level. The core type, on the other hand, is the best choice for greater operating frequency [5]. To increase the efficiency of the transformer it is high time to focus on the core materials property and structure of the transformer.

### 2.1 Geometry of U-I Shaped Lamination

The magnetic core of a core-type transformer is made up of laminations that form a rectangle-shaped frame. However, the entire amount of iron as well as the volume might be too enormous. Pooling a core enables decreasing the overall mass of iron. In such a balanced system, overall flow through all the core iron becomes zero as that's the composition from three balanced flows. A single-phase core type involves a larger magnetic core with two wound limbs(fig. 1).

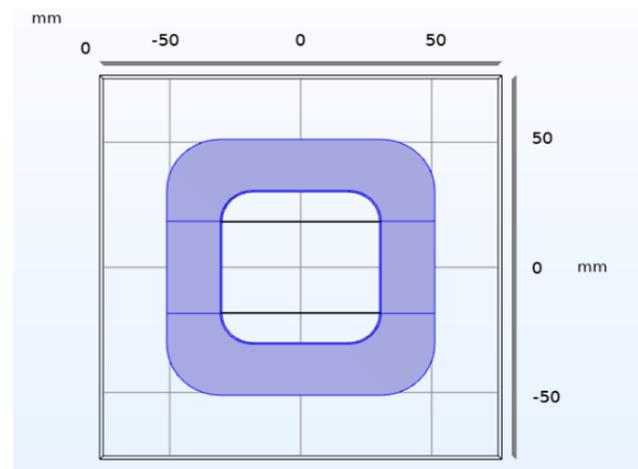

**Fig. 1:** Geometry of U-I shaped Core type transformer.

### 2.2 Geometry of E-I Shaped Lamination

A three-limb core is used to produce the core of a single-phase shell type transformer. This design improves the core's mechanical strength. It also improves the safety of windings against mechanical shocks from outside sources. In fig. 2 a single-phase shell type contains two magnetic E-cores which are coupled in a manner to construct a three-limb transformer, the center limb has been wound.

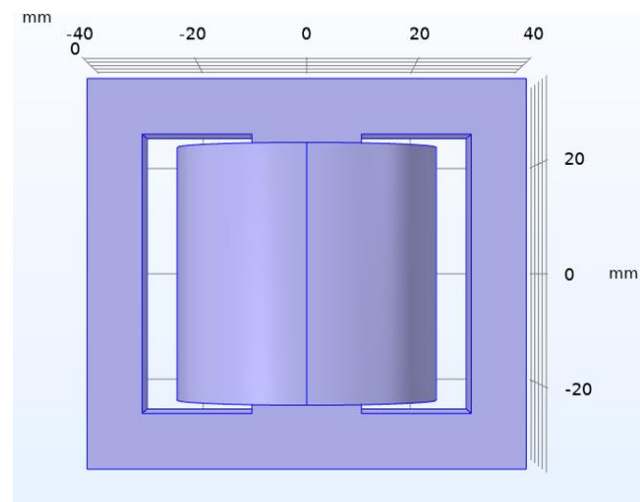

**Fig. 2:** Geometry of E-I shaped Shell type transformer.

### 3. MODELING AND SIMULATION PROCESS

#### 3.1 Methodology

For modeling the transformer, the commercial finite element program COMSOL Multiphysics® Version 5.6 is used. In engineering and mathematics, the finite element method (FEM) seems to be an extensively accepted methodology for numerically solving differential equations. The FEM is indeed a basic numerical strategy for solving partial differential equations which take into account two or three spatial variables. This is completed by creating a mesh of the object: a numerical domain for the solution, with a finite number of points, that would be accomplished via a specific space finite difference method within spatial dimensions.

#### 3.2 Input Materials

Materials utilized throughout this modeling would be those that help in a transformer; hence, we obtain three main materials i.e: Iron and Ferrite.

**Soft Iron.** The iron is being used as just a ferromagnetic material for simulating the transformer, whose primary function is to network magnetic flux. Table 1 shows the properties of iron.

**Ferrite.** The properties of Ferrite are shown in Table 1. Ferrite is a material having magnetic characteristics which may be used in a variety of electrical devices. Ferrite is also used to channel the magnetic flux in the transformer, used for modeling the ferromagnetic transformer core.

**Table 1.** Properties of Soft Iron and Ferrite.

| Material property | Soft Iron | Ferrite |
|---|---|---|
| Relative permeability | 2,00,000 | 4,000 |
| Electrical Conductivity | 1(S/m) | 1(S/m) |
| Relative permittivity | 1 | 1 |

#### 3.3 Parameters and Variables

A parameter estimation study in COMSOL Multiphysics® can be used to estimate the values of inputs in a model. Parameter nodes are found under Global Definitions and can be used to create and define global parameters. There is always one parameter node accessible, and it cannot be moved or deleted. When arranging the parameters in numerous nodes, possibly with descriptive labeling, further parameter nodes can be added as needed.

**Table 2.** Parameters for U-I shaped laminated transformer.

| Name | Description | Value |
|---|---|---|
| f | Frequency | 60 Hz |
| Ns | Number of turns in secondary windings | 100 |
| Np | Number of turns in secondary windings | 20 |

**Table 3.** Parameters for E-I shaped laminated transformer.

| Name | Description | Value |
|---|---|---|
| Rp | Primary windings resistance | 100 ohm |
| Ns | Number of turns in secondary windings | 300 |
| f | Frequency | 50 Hz |

To specify expressions as user-defined variables, use the Variables node. Global variables can be utilized in any context that takes variable expressions, including all Components and geometric entities, as long as their expressions are likewise global. Local variables, on the other hand, have a defined geometric domain. Such variables can only be utilized and assessed within a single Component, or on a limited number of domains, boundaries, edges, or points.

**Table 4.** Variables for U-I shaped laminated transformer.

| Name | Description | Value |
|---|---|---|
| a | Steinmetz constant | 50 |
| b | Steinmetz constant | 1.6 |
| c | Steinmetz constant | 1 |
| PL | Power loss | $a*(mf.normB)^b*(f)^c$ |
| PL_total | Total power loss | intop1(PL) |

**Table 5.** Variables for E-I shaped laminated transformer.

| Name | Description | Value |
| --- | --- | --- |
| Rs | Secondary side windings | 10 ohm |
| Np | Number of turns in the primary winding | 300 |
| V | Supply voltage | 25 V |

### 3.3 Mesh

Any numerical solution to a problem, particularly one using finite elements, must begin with the design of a mesh depicting the field of inquiry. This stage is frequently delicate since the accuracy of finite element approximation is heavily dependent mostly on mesh structure. In the mesh analysis maximum element size is 5.25 mm and minimum element size is 0.225 mm from one node to another. Fig. 3 and Fig. 4 represents extremely fine meshing in U-I and E-I shaped transformer.

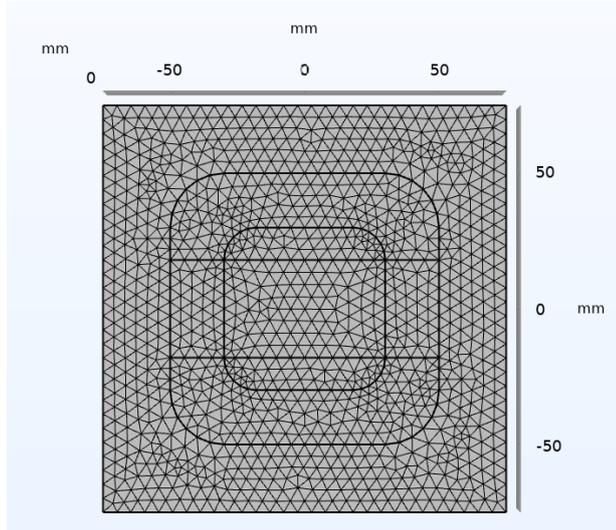

**Fig. 3:** Meshing of U-I shaped Shell type transformer.

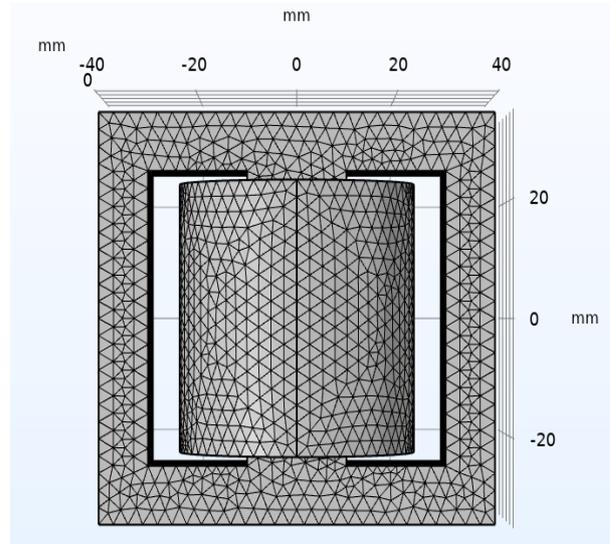

**Fig. 4:** Meshing of E-I shaped Shell type transformer.

### 4. SIMULATION RESULT

**4.1 Magnetic Flux Density**

Fig. 5 is the magnetic flux density profile of the E-I shaped transformer whose core part is formed by ferromagnetic soft iron. This profile is calculated by simulation software by studying the frequency domain step. Magnetic flux density(B) for ferromagnetic soft iron core's transformer has a quite uniform profile. Here, B is distributed in the core from 0.1 T to 0.6 T.

As E-I shaped laminated transformer has a lower air gap, it shows lower permeability($\mu$) which means a lower value of B than U-I shaped transformer. The reason behind this is the magnetic flux density is proportional to the permeability of the material. But shows a higher value of B than the core of the Ferrite as soft iron's value of $\mu$ is higher than the Ferrite.

In the Density of the magnetic flux profile, the magnetic field value is always at the saturation elbow. This could be owing to the presence of two additional flux return columns, which allow the magnetic field lines in the ferromagnetic circuit to travel freely.

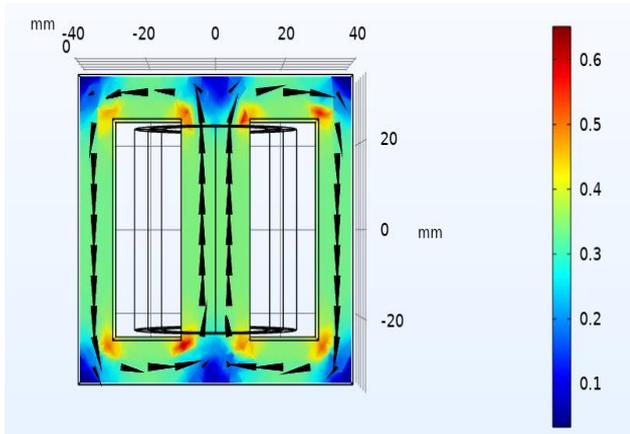

**Fig. 5:** Magnetic Flux Density of E-I shaped transformer (Soft Iron Core).

Fig. 6 illustrates the magnetic flux density profile of the E-I shaped transformer whose core part is formed by ferrite. Here, B is distributed in the core from 0.05 Tesla to 0.3 Tesla. However, When Ferrite is used in the E-I shaped transformer, B values are lower than the Soft Iron's core. As ferrite's µ is lower than the soft iron. Because B is proportional to the µ of the material, values of B of the ferrite core are lower than the soft Iron Core.

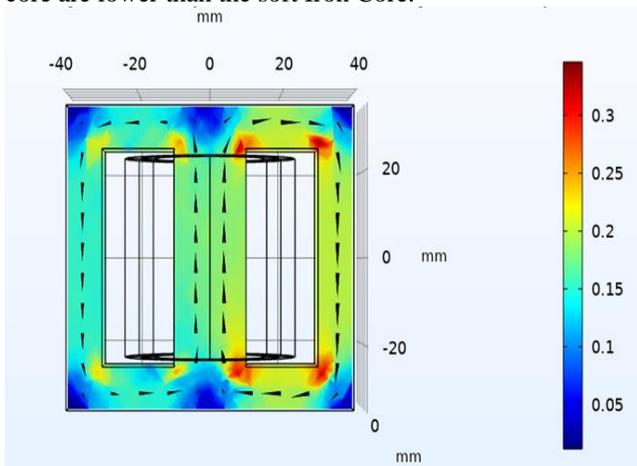

**Fig. 6:** Magnetic Flux Density of E-I shaped transformer (Ferrite Core).

Fig. 7 is the magnetic flux density profile of the U-I-shaped transformer whose core part is formed by ferromagnetic soft iron. This profile is calculated by simulation software by studying the frequency domain step. Here, B is distributed in the core from 7.08 ×10$^{-8}$ T to 21.1 T. Number of windings of the secondary coil in this transformer is higher than the primary coil. As Number of secondary coils windings and primary coils windings is 100 and 20. B is proportional to the number of windings. From this profile, higher values of B were found near the secondary coils. As U-I shaped laminated transformer has a higher air gap, it shows higher values of µ which means a higher B value than E-I shaped laminated transformer. As soft iron has a higher value of µ than Ferrite. B value is higher in the soft iron core of the U-I laminated transformer than that of the Ferrite core because B is proportional to the material's µ.

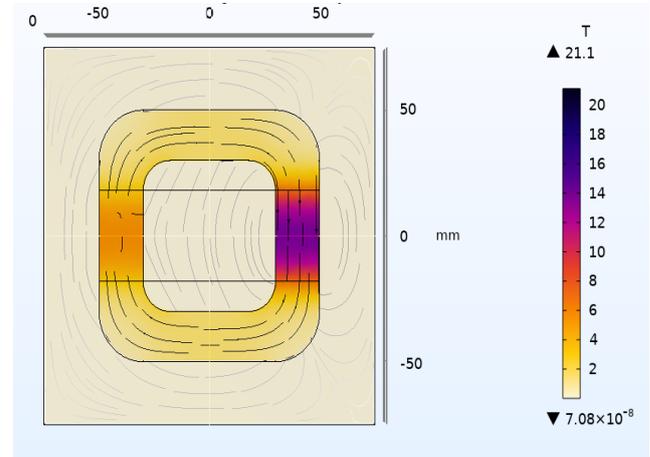

**Fig. 7:** Magnetic Flux Density of U-I shaped transformer (Soft Iron Core).

Fig. 8. is the magnetic flux density profile of the U-I shaped transformer whose core part is formed by ferrite. Here, B is distributed in the core from 7.4 ×10$^{-8}$ T to 14.1 T. However, When Ferrite is used in the U-I shaped transformer, B values are lower than the Soft Iron's core. Because ferrite has a lesser µ than soft iron. B values of the ferrite core are lower than that of the soft iron core because B ∝ µ.

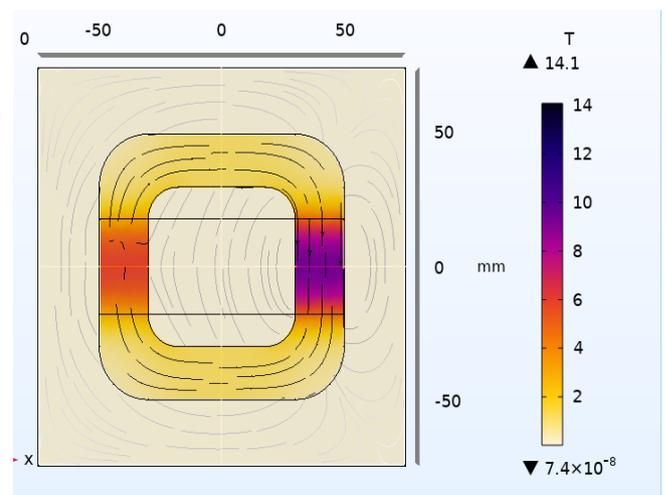

**Fig. 8:** Magnetic Flux Density of U-I shaped transformer (Ferrite Core).

**4.2. Volumetric loss density**

Fig. 9 is the volumetric loss density profile of the E-I shaped transformer whose core part is formed by ferromagnetic soft iron. Here, volumetric loss density is distributed in the core from 86.3 W/m$^3$ to 3.48×10$^4$ W/m$^3$. Volumetric loss density for ferromagnetic soft iron core's transformer has a quite uniform profile but values are moderately higher.

Multiplying the volumetric loss density by the effective core volume generates a total core loss. As the volumetric loss density rises, the total core loss rises with it. Core loss is more due to total flux through the entire core. When the air gap of the transformer is increased, reducing the slope of the B/H loop, reducing inductance and μ. The air gap also prevents core saturation. Less hysteresis loss is represented by a smaller hysteresis loop area (B/H loop). For lower air gap, the loss is higher comparatively in E-I shaped transformer than U-I shaped laminated transformer.

Because the core of a transformer is formed of soft iron magnetizing material, when a current passes through it, an alternating magnetic field is created, which causes a little current to flow through the core material again, increasing core losses.

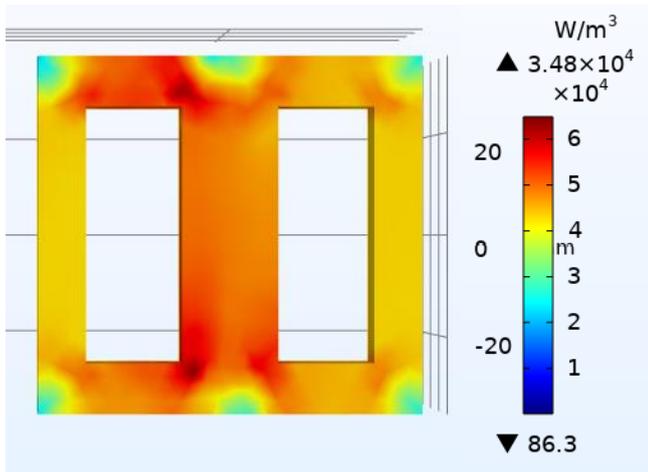

**Fig. 9:** Volumetric loss Density of E-I shaped transformer (Soft Iron Core).

Fig. 10 is the volumetric loss density profile of the E-I shaped transformer whose core part is formed by ferrite. Here, volumetric loss densities are distributed in the core from 63.2 W/m$^3$ to 3.26×10$^4$ W/m$^3$. Volumetric loss density for ferrite core's transformer has also quite uniform profile but values are moderately lower than soft iron's cored E-I transformer.

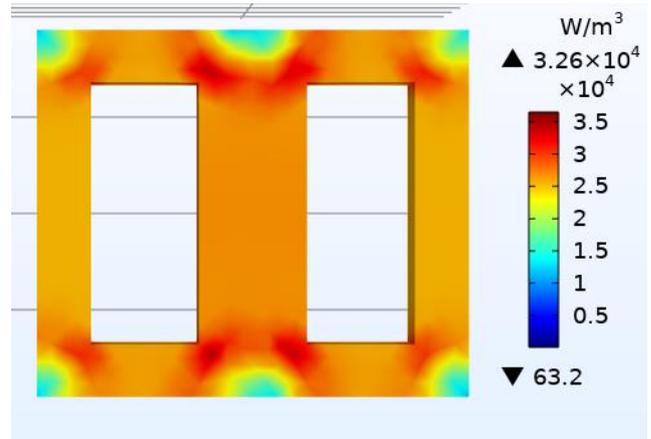

**Fig. 10:** Volumetric loss Density of E-I shaped transformer (Ferrite Core).

Fig. 11 is shown the Volumetric loss Density of U-I shaped transformer where Soft Iron is used as the Core material. Total core loss is attained by multiplying the volumetric loss density by the effective core volume. As, with increasing Volumetric loss density, total core loss is also increased. The secondary coil loss profile shows a higher volumetric loss density. Other parts of the core loss are in the range from $3.88 \times 10^{-6}$ w/m$^3$ to 1 w/m$^3$. One reason for having core loss is lower in U-I lamination type transformer is using frequency above 5 kHz as the model's parameter. When the transformer's air gap is raised, the slope of the B/H loop is reduced, lowering permeability and inductance. The air gap also keeps the core from becoming saturated. A smaller hysteresis loop area (B/H loop) indicates less hysteresis loss. So, another reason, for larger air gap loss is less comparatively in U-I shaped transformers than E-I shaped laminated transformers. This is because only half of the total flux flows in the core because of having a large air gap.

As the transformer core of U-I lamination is made from soft iron magnetizing material, whenever a current flow around then an alternating magnetic field is produced that inures again small current to flow in the core material and increased the core losses than Ferrite transformer core.

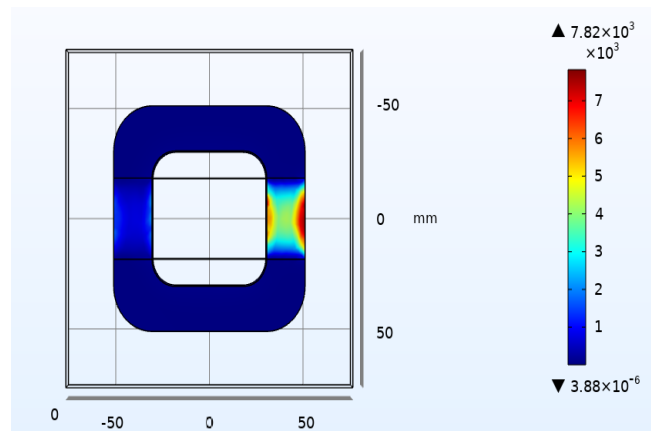

**Fig. 11:** Volumetric loss Density of U-I shaped transformer (Soft Iron Core).

From fig. 12 Volumetric loss Density has been found for U-I shaped transformers where Ferrite is used as Core material. The secondary coil loss profile shows a higher volumetric loss density. Other parts of the core loss are in the range from $1.41 \times 10^{-7}$ w/m$^3$ to 1 w/m$^3$. However, When Ferrite is used in the U-I shaped transformer, Volumetric loss density values are lower than the Soft Iron's core.

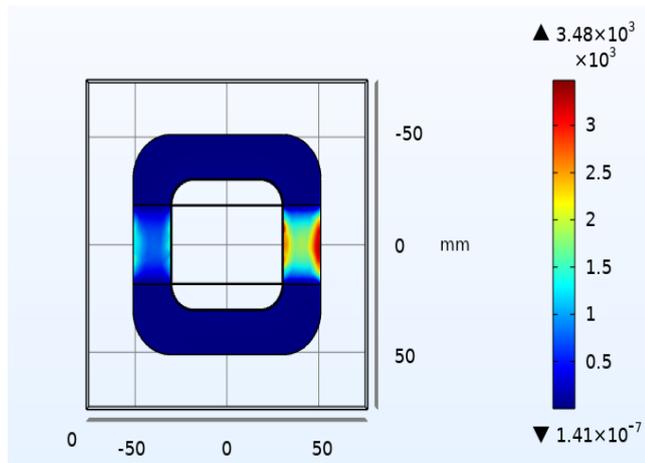

**Fig. 12 :** Volumetric loss Density of U-I shaped transformer (Ferrite Core).

## 4. CONCLUSION

In this research different structure transformers have been modeled through COMSOL Multiphysics simulation software and COMSOL-based FEM electromagnetic field and magnetic properties of them are analyzed. The findings of simulations have aided in determining the impact of various materials on those transformers. Magnetic flux density and volumetric loss density profile are plotted. The core material is varied and its impact on material properties is investigated. Supreme values of magnetic flux density are obtained in the Core Type (U-I laminated) transformer. Volumetric loss density is found higher in the shell type (E-I laminated) transformer. Magnetic flux density and volumetric loss density are both higher with using soft iron as core material than Ferrite core. Core type transformer is simulated at a higher operating frequency(60 Hz) than shall type transformer(50 Hz). The air gap should be higher for having less Volumetric loss in the transformer. Machine cores become easier to maintain as a result of fewer core loss of core type (U-I laminated) ferrite materials core. According to the magnetic flux density and volumetric loss density, the most significant and determining factors of the E-I shell type transformer is the excellent solution in the low range of operating frequency. The U-I core type, on the other hand, is the best choice for greater operating frequencies. U-I type transformer is most feasible for 3D printing.

In this analysis, only magnetic flux density and volumetric loss density properties were investigated only a limited range. In the future using other ferromagnetic material (Mu-metal,permalloy-78) hysteresis loss curve, current density distribution, heat transfer, the temperature distribution of L-L lamination, E-E lamination and other structured transformers can be analyzed. Transformer cores will be 3D printed in the future with FDM 3D printers. Property investigation of the 3d printed transformer will be continued with the practical investigation to verify the simulated result.

This is a notable fact that transformer topology is characterized by a significant amount of core loss, which may be a serious problem for some applications despite its strong potential for efficient magnetic density and volumetric loss density.

## REFERENCES

[1] N. Kitaev, Ł. Kaiser, and A. Levskaya, "Reformer: The Efficient Transformer," pp. 1–12, 2020, [Online]. Available: http://arxiv.org/abs/2001.04451

[2] G. Bertotti, "Physical interpretation of eddy current losses in ferromagnetic materials. II. Analysis of experimental results," *J. Appl. Phys.*, vol. 57, no. 6, pp. 2118–2126, 1985, doi: 10.1063/1.334405.

[3] T. Q. Pham, T. T. Do, P. Kwon, and S. N. Foster, "Additive Manufacturing of High Performance Ferromagnetic Materials," *2018 IEEE Energy Convers. Congr. Expo. ECCE 2018*, no. January 2019, pp. 4303–4308, 2018, doi: 10.1109/ECCE.2018.8558245.

[4] K. Onla, C. Junior, K. Fankem, E. Duckler, D. T. Alix, and E. J. Yves, "3D Multiphysics Modelling of Three-Phase Transformer," vol. 10, no. 2, pp. 33–45, 2021.

[5] M. A. Bahmani, E. Agheb, T. Thiringer, H. K. Hoidalen, and Y. Serdyuk, "Core loss behavior in high frequency high power transformers-I: Effect of core topology," *J. Renew. Sustain. Energy*, vol. 4, no. 3, 2012, doi: 10.1063/1.4727910.

[6] F. Edition and C. April, "37, 311 (1951).," vol. 39, no. 1951, pp. 551–560, 1953.

[7] P. Pfwvdl *et al.*, "Characteristic Comparison of Transv versally Laminated Aniso otropic Synchronous Reluctance Motor Fabricatio on Based on 2D Lamin ation and 3D Printing," pp. 18–21.

[8] C. Ropoteanu, P. Svasta, and C. Ionescu, "Electro-thermal simulation study of different core shape planar transformer," *2016 IEEE 22nd Int. Symp. Des. Technol. Electron. Packag. SIITME 2016*, pp. 209–212, 2016, doi: 10.1109/SIITME.2016.7777279.

[9] K. S. Shinoy and B. Sebastian, "Modeling a Brushless DC Motor for an Advanced Actuation System Using COMSOL Multiphysics® Software," *Comsol.Jp*, [Online]. Available: https://www.comsol.jp/paper/download/367791/ks_abstract.pdf.

[10] G. Kron, "Generalized Theory of Electrical Machinery," *Trans. Am. Inst. Electr. Eng.*, vol. 49, no. 2, pp. 666–665, 1930, doi: 10.1109/T-AIEE.1930.5055554.